\documentclass[]{article}
\usepackage{amsmath, graphicx}

\begin{document}

\title{Bend-Free Multiarm Interferometers on Optical Chips}


\author{\small{Jovana Petrovic$^{1,*}$, Aleksandra Maluckov$^1$, Nikola Stojanovic$^2$}\\
\small{\textit{$^{(1)}$ Vinca Institute of Nuclear Sciences, National Institute of Republic of Serbia,}}\\ \small{\textit{ University of Belgrade, Mike Petrovica Alasa 12-14, 11000 Belgrade, Serbia}}\\
\small{\textit{$^{(2)}$ Deutsches Zentrum f\"{u}r Luft- und Raumfahrt (DLR),}}\\\small{\textit{Rutherfordstra{\ss}e 2, 12489 Berlin, Germany}}\\
\small{\textit{$^*$E-mail: jovanap@vin.bg.ac.rs}}}

\date{}
\maketitle

\begin{abstract}
Multiarm interferometers can enhance measurement precision and provide multiparameter capability to the measurement. Their realisation requires multiport beam splitters, which has been a long-standing challenge in free-space and integrated optics. Here, we propose a new type of multiport interferometers suitable for implementation on optical chips. Their prospective advantages over the standard directional-coupler architectures: an arbitrary number of arms, planar architecture and two orders of magnitude reduction in footprint, are achieved by the layout based exclusively on the finite modulated photonic lattices. The inverse design of photonic lattices is facilitated by restricting the light propagation to periodic patterns. The corresponding interferometer model predicts the maximum $1/\sqrt{N}$ sensitivity scaling with the number of arms $N$. While the presented design solutions are ubiquitous to all implementations, the advantages are discussed in both the low and high refractive-index contrast platforms. Finally, we provide an outlook on the possibilities for the interferometer applications in distributed classical and quantum measurements.
\end{abstract}

\section{Introduction}\label{sec:Introduction}
Interferometry is the preferred technique for measurements of small phase changes. The best achievable sensitivity of a classical two-path interferometer is determined by the shot noise and is given by $1/\sqrt{P}$ scaling with the number of measurements $P$, known as the standard quantum limit (SQL). The ultimate sensitivity enhancement entails the use of entangled states and a collective non-local measurement at the output. Thus achievable sensitivity scaling $1/P$ is known as Heisenberg limit \cite{GiovannettiScience2004}. An alternative way to improve the measurement outcome is the multiplexing of interferometer modes which can be seen as a super-resolving technique~\cite{WeitzPRL1996, PetrovicNJP2013, SuPRL2017}. For the phase sensing by $N<P$ uncorrelated interferometric modes, the maximum sensitivity enhancement over the shot-noise limit is $\sqrt{P/N}$~\cite{PezzeNatPhoton2021}. Besides the opportunities for the resolution enhancement, the multimode interferometers enable multiparameter measurements, which are on high demand in structural health, medical and environmental monitoring~\cite{ChennanOptExpress2016, SunJRSM2020}.

Realisation of multiarm interferometers requires multiport beam splitters and combiners, which have been a long-standing challenge in free-space optics. Development of optical integrated and fibre technology has enabled construction of multiports \cite{SuzukiOptExress2006} and hence multiarm interferometers ~\cite{WeihsOptLett1996,ChaboyerSciRep2015}. The integrated multiports are realised as directional couplers with arms expanding in two or three dimensions~\cite{KeilAPLPhoton2016, SpagnoloSciRep2012, ChaboyerSciRep2015}. The principal feature of a directional coupler is the proximity of waveguides in its central region which allows for efficient transfer of energy between them. To enable the input and output light coupling into individual waveguides, the waveguides fan out from the central region towards the input and output ports. The bending radii associated with fanning are limited from below by the radiation losses \cite{MarcatiliBellSysTechJ1969}. This makes the extension to a larger number of arms challengeable. Moreover, branching in all three dimensions results in different depths of the interferometer arms, which requires an intricate phase control in the sensing region. While nesting of  2x2 directional couplers provides multiple arms in planar geometry and thus circumvents this problem \cite{PolinoOptica2019}, it requires repeated bends. Repeated bends lead to higher losses or an increase in the circuit footprint.

We propose photonic-lattice (PL) based multiarm interferometers that are planar, bend-free and with the footprint limited only by the crosstalk between waveguides. In the paper, we first present the interferometer layout, describe in detail numerical design of its multiport couplers and the rationale for the phase distribution in the sensing region. We then demonstrate the fringe narrowing with the number of arms - the key signature the multiarm interferometry and show that the sensitivity scaling complies with the established knowledge. Finally, we discuss the achievable reduction in the interferometer footprint, disadvantages and possible advantages of the crosstalk in the sensing region and multiparameter sensing.

\section{Interferometer design}\label{sec:Interferometers}
\begin{figure}[htb]%
\centering
\includegraphics[width=0.9\textwidth]{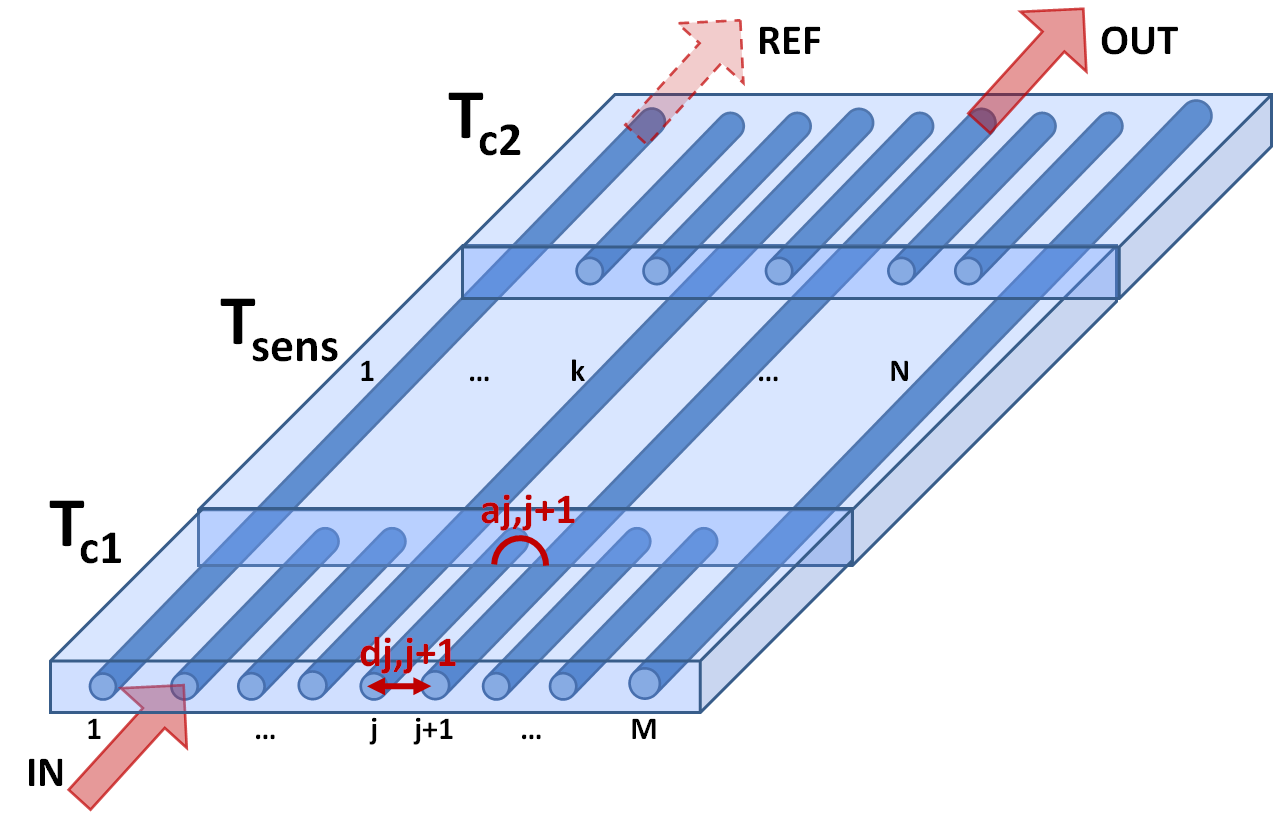}
\caption{Layout of an unfolded $N$-arm Michelson interferometer on chip. It is composed of a beam splitter realised as $1\times N$ coupler $T_{c1}$, sensing region $T_{sens}$ and beam combiner realised as $N\times1$ coupler $T_{c2}$. The sensing region is engineered to admit a linear phase change across waveguides when exposed to a single external agent. The parameters $a_{j,j+1}$ and $d_{j,j+1}$  denote the coupling coefficients between the neighboring waveguides and the corresponding distances, respectively.}\label{fig:Scheme}
\end{figure}
The interferometer is composed of three PLs. Two PLs comprise linearly coupled waveguides and act as beam splitters. They sandwich the third PL with a negligible coupling between waveguides that serves as a sensing element upon which an external parameters acts, Fig.~\ref{fig:Scheme}.
The constitute PLs are described by the respective transfer matrices $\mathbf{T}_{c1}$, $\mathbf{T}_{c2}$ and $\mathbf{T}_{sens}$. The output of the interferometer is given by $\psi(z)=\mathbf{T}\psi(z=0)$, where $\mathbf{T}=\mathbf{T}_{c2}\mathbf{T}_{sens}\mathbf{T}_{c1}$ is the transfer matrix of the interferometer and $\psi(z)=(\psi_1(z), \psi_2(z),\ldots,\psi_j(z),\ldots,\psi_M(z))$ the state vector of the mode wavefuntions $\psi_j(z), j=1,\ldots M$. The measurement observables are the light intensities $I_j=\psi_j\psi^*_j$ measured at all output ports $j=1,\ldots M$ simultaneously.

In the tight-binding model, the PL transfer matrix $\mathbf{T}$ is obtained by solving the Schr\"{o}dinger equation
\begin{equation}
i\frac{d\psi(z)}{d z}= \mathbf{H}\psi(z),
\label{eq:Model}
\end{equation}
where the PL Hamiltonian is represented by a tridiagonal matrix
\begin{equation}\label{eq:H}
{H\mathbf{}}=\left(
\begin{array}{ccccc}
\Delta_{1,1} & a_{1,2} &&&\\
a_{2,1} & & &  &  \\
 &\ddots & \ddots& \ddots&\\
 & & & &  a_{M,M-1}\\
& & &  a_{M-1,M} &\Delta_{M,M}\\
\end{array}
\right).
\end{equation}
Its off-diagonal terms $a_{k,k\pm1}$ are the $z$-invariant coupling coefficients between adjacent waveguides, Fig. \ref{fig:Scheme}. Its diagonal terms $\Delta_{k,k}$ model the phase mismatch between the modes of different waveguides \cite{BrombergPRL2009}. Integration of (\ref{eq:Model}) yields $\psi_k(z)=\sum_{j=1}^M{\mathbf{T}_{k,j}(z)\psi_j(0)}$, where $\mathbf{T}_{k,j}(z)=(e^{-i\mathbf{H}z})_{k,j}$ constitute the unitary transfer matrix $\mathbf{T}$. The nearest-neighbour approximation is justified by the exponential decay of the coupling coefficients with the distance between waveguides \cite{BellecOptLett2012, GuzmanPRL2021}. This has been corroborated by numerical simulations \cite{PetrovicAnnPhys2018}.

\subsection*{Couplers}
The key task in the multiport-coupler design is finding the PL Hamiltonian from the known output and input states. This is the inverse problem that reaches high complexity for large number of ports. Here, we consider two types of couplers that produce binomial or equal intensity distributions at the output. The former are intrinsic to the atom interferometers and are well-described by Clebsch-Gordan coupling coefficients \cite{PetrovicNJP2013}. Their photonic analogues have been exploited in realisation of the faithful information transfer in photonic lattices \cite{BellecOptLett2012}. The latter, albeit used in a vast majority of interferometric schemes, are not trivial to design in a discrete space. However, the recent interest in multipartite-entangled W-states, that are characterised by equal probability of all modes, has accelerated the efforts and a number of particular solutions have been found numerically~\cite{Perez-LeijaPRA2013,PaulJOpt2014,GraefeNatPhoton2014}.

Taking the analytical approach instead, we have shown that by instigating the periodicity into the light propagation through a PL, the inverse problem can be reduced to a few-paramater optimisation problem, thus permitting a fairly simple design of $1\times N$ couplers \cite{PetrovicOptLett2015}. This is achieved by imposing the commensurability of all eigenvalues of the PL Hamiltonian.

We demonstrate the procedure on the couplers and interferometers shown composed of symmetric and asymmetric PLs with 5 waveguides. Their coupling coefficients are expressed as functions of the arbitrary unequal integers $n_1$ and $n_2$. A detailed derivation is given in~\cite{PetrovicAnnPhys2018} and results in
\begin{equation}
\begin{array}{ccl}\label{eq:SolutionM5asymm}
a_{1,2} & = & 1\\[0.0cm]
a_{2,3} & = & \sqrt{
\frac{(n_1^2n_2^2-s^2t^2)-(n_1^2+n_2^2-(s^2+t^2))s^2}{s^2(t^2-s^2)}}\\[0.3cm]
a_{3,4} & = & \sqrt{
\frac{-(n_1^2n_2^2-s^2t^2)+(n_1^2+n_2^2-(s^2+t^2))t^2}{s^2(t^2-s^2)}}\\[0.3cm]
a_{4,5} & = & \frac{t}{s}
\end{array}
\end{equation}
for an array without symmetry around the central waveguide. Here, 0, $\pm n_1$ and $\pm n_2$ are the eigenvalues of the lattice Hamiltonian and $s$ and $t$ are any reals that render real coupling coefficients. The mirror-symmetric lattices are defined by
\begin{equation}
\begin{array}{ccl}
a_{1,2} = a_{4,5} & = & 1\\[0.2cm]
a_{2,3} = a_{2,3} & = & \sqrt{\frac{n_2^2-n_1^2}{2n_1^2}}.
\end{array}
\label{eq:SolutionM5symm}
\end{equation}
The Hamiltonian of a symmetric PL and hence the coupling ratios that it can support are fully determined by its eigenvalues. On the other hand, the Hamiltonian of an asymmetric PL depends on two additional parameters $s$ and $t$ that can be used to optimize the coupling ratio. Here, we fix the eigenspectrum and search for $1\times N,\,N=2,3,4,5$ splitters by exploring the three-dimensional $\{s,t, L\}$ parameter space, where $L$ is the lattice length. Remarkably, it is possible to find the parameters that render the wanted coupling ratio with an arbitrary small error. For practical purposes, we set the maximum discrepancy from the wanted intensity at any output port to 0.2\%.

\subsection*{Phase distribution in the sensing region}
The profile of the phase-change across the interferometer arms determines the fringe shape. Integrated optics offers unprecedented possibilities to control the impact of a single parameter on different waveguides and thus shape the phase-change profile. We follow the Fourier approach and build a linear phase profile with a constant phase mismatch between waveguides. This renders the transfer matrix $\mathbf{T}^{sens}_{j,k}=e^{-i(j\Delta\Phi)\delta_{j,k}}$, where $\Delta\Phi$ is the phase change in the fist arm, $j\Delta\Phi$ the phase change in $j^{th}$ arm and $\delta_{j,k}$ the Kronecker symbol. This concept has been corroborated in a free-space optical quantum Fourier interferometer \cite{SuPRL2017}, as well as an atomic multi-state interferometer \cite{PetrovicNJP2013}. In an optical fiber or integrated interferometer, a linear phase distribution can be achieved by a temperature gradient across the chip \cite{LedererPRL1999}, insertion of optofluidic channels of different lengths \cite{PetrovicAO2008, CrespiLOC2010}, integrated electro-optical modulators \cite{ZhangOptica2021}, etc.

\section{Results}\label{sec:Results}
\begin{figure}[ht]%
\centering
\includegraphics[width=0.9\textwidth]{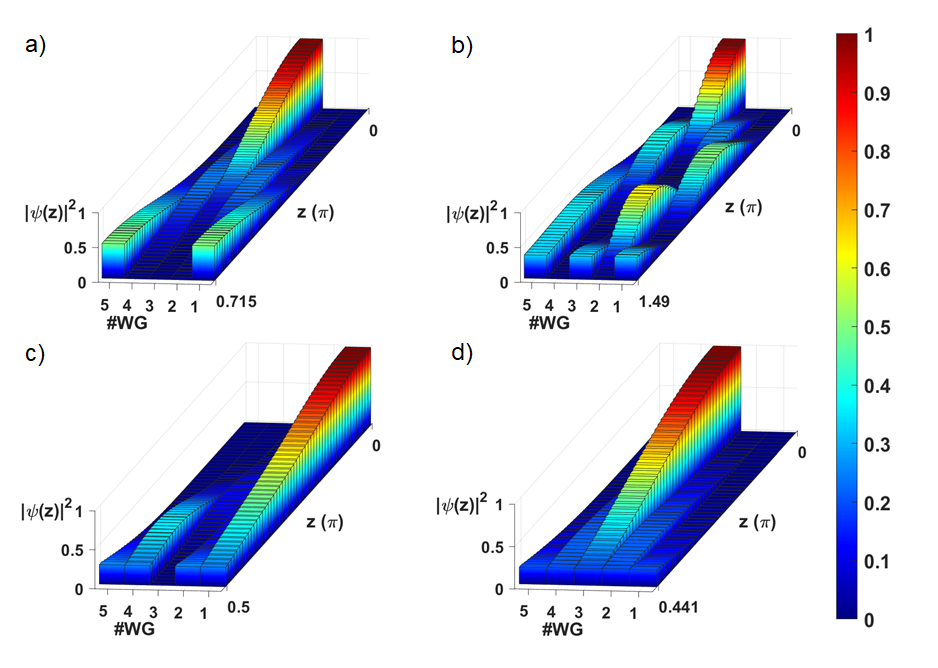}
\caption{Inverse solutions for $1\times N$, $N=2,3,4,5$ couplers based on 5-waveguide commensurable photonic lattices. Shown is the intensity pattern of the propagation along the PLs (z-axis). The total intensity of the input state is normalized to 1. The coupling coefficients are calculated using equations (\ref{eq:SolutionM5asymm}) and (\ref{eq:SolutionM5symm}). a) $1\times 2$ coupler with the the coupling coefficients $a_{1,2}={4,5}=1$ and $a_{2,3}=a_{3,4}=0.6928$ b) $1\times 3$ coupler with  $a_{1,2}=1$, $a_{2,3}=0.7495$, $a_{3,4}=0.689$ and $a_{4,5}=0.4644$, c) $1\times 4$ coupler with $a_{1,2}=a_{4,5}=1$ and $a_{2,3}=a_{3,4}=2.7386$ and d) $1\times 5$ coupler the coupling coefficients $a_{1,2}=a_{4,5}=1$ and $a_{2,3}=a_{3,4}=0.6236$. Note that $1\times 3$ coupler is an asymmetric PL while the others are symmetric.}\label{fig:Couplers}
\end{figure}

Results in Fig.~\ref{fig:Couplers} show that our method can be used to construct $1\times N$ equal-power splitters with an arbitrary $N$ and choice of output ports. The latter is particularly important for construction of multiarm interferometers that require the minimum possible coupling between waveguides in the sensing region and hence significant separations between the output ports. For example, the couplers in a) and b) accommodate this spacing, while the couplers in c) and d) maintain the coupling between waveguides throughout the sensing region. Indeed, the asymmetric PL in c) has been optimized to avoid adjacent output ports, whereby we assume that the next-to-nearest-neighbour coupling is negligible with respect to the nearest-neighbour coupling. The assumption is based on the exponential dependence of the coupling coefficient on the waveguide distance and corroborated numerically. The simulations have shown that the coupling beyond the next neighbours can be treated as a perturbation to the eigenspectrum and consequently to the periodic light dynamics \cite{PetrovicOptLett2015}. As the effects of this perturbation become significant only upon several propagation periods, they are negligible on the length scale of the interferometers considered here.

\begin{figure}[ht]%
\centering
\includegraphics[width=\textwidth]{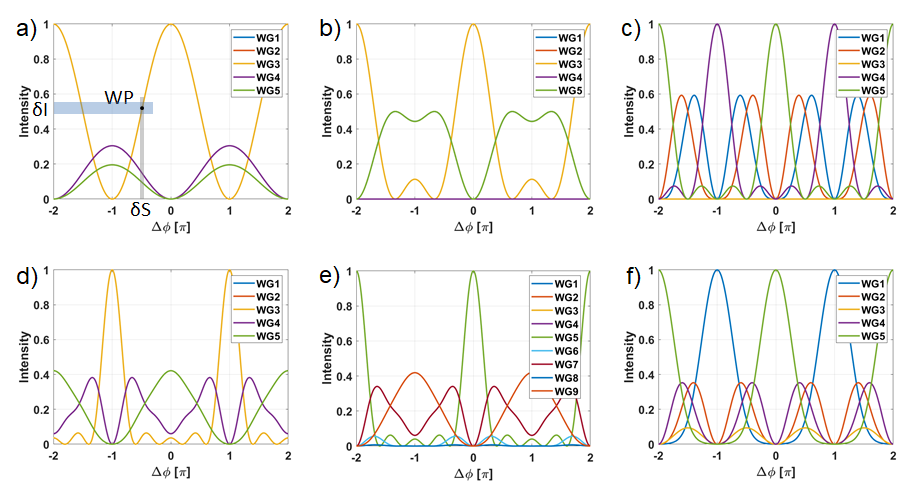}
\caption{a)-d) Fringe patterns of interferometers built using the couplers from figure \ref{fig:Couplers} a)-d), respectively, under the assumption that the waveguides do not interact in the sensing region. \textbf{The total intensity of the input state is normalized to 1.} e) 5-arm interferometer constructed with the $1\times5$ couplers with the light distributed equally between the next-to-nearest ports (1,3,5,7,9). The coupler is based on the 9-waveguide array with the coupling coefficients $a_{1,2}=a_{8,9}=4.029$, $a_{2,3}=a_{7,8}=2.482$, $a_{3,4}=a_{6,7}=2.615$, $a_{4,5}=a_{5,6}=1$ found in \cite{PaulJOpt2014}. f) 5-arm interferometer with the couplers that exhibit spin-like coupling of waveguides with Clebsch-Gordan coupling coefficients $a_{1,2}=a_{4,5}=1$, $a_{2,3}=a_{3,4}=\sqrt{3/2}$.}\label{fig:Int}
\end{figure}

Impact of multiple arms on the measurement uncertainty depends on the choice of observable. The interferometer in Fig.~\ref{fig:Scheme} is designed to probe a local phase change around the working point (WP in in Fig.~\ref{fig:Scheme} a)) optimally placed at the maximum slope. Tuning of the working point can be realised by a phase changing mechanism independent of the measured parameter, for example, an interferometer that works as a temperature sensor can be tuned by an electro-optic effect. The sensitivity $\delta S$ is defined as the minimum phase change that the interferometer can detect and is estimated as $\delta S=\big(\frac{\delta I}{\partial I/\partial \Delta\phi}\big)_{min}$, where $\delta I$ is the minimum signal detectable above the noise and $\partial I/\partial \Delta\phi$ the fringe slope. The split of photons into arm causes decrease in the average number of photons per arm and thus greater susceptibility to noise. Assuming the shot-noise limit, the minimum detectable intensity increases by $\sqrt{N}$. On the other hand, the fringe slope scales linearly with $N$. This is well-known result for ideal multiarm interferometer, here confirmed for interferometers in Fig.~\ref{fig:Int} a)-e). As a consequence, the sensitivity improves by $\sim 1/\sqrt{N}$.

The interferometer sensitivity depends on the splitting (coupling) ratio of its couplers. For example, the couplers with a binomial ratio of intensities at the output ports (1:4:6:4:1) used in Fig.~\ref{fig:Int} f) render wider fringes with a smaller sensitivity  than the equal-power splitters. We have investigated other splitting ratios and have not found any that gives better sensitivity than the equal-power splitter, the result well known in the two-path interferometry. The interferometer signal does not depend on the relative phases at the splitter outputs, but solely on the coupling ratio. For example, the interferometers in d) and e) are constructed with $1\times 5$ couplers realised with 5- and 9-waveguide lattices, respectively. The phases at the outputs of the former  $(0,\,-\pi/2,\,0,\,-\pi/2,\,0)$ and of the latter $(0,\, 0,\, 0,\, 0,\, 0)$ are markedly different, yet render the same interferometer sensitivity.

We note that a similar analysis applies to the interferometric measurements designed to detect a shift in fringes. Here, the fringe width determines the measurement resolution. The width of the narrowest fringe of the interferometers in Fig.~\ref{fig:Int} a)-d) is inversely proportional to the number of arms. The same scaling was obtained for two definitions of the fringe width: the full-width half-maximum and the width between the two side nodes. As above, the immense fringe narrowing at a very large number of arms is counteracted by the reduction in the number of photons per arm, which prevents the infinitesimal sensitivity.

\section{Discussion}\label{sec:Discussion}

Elimination of the waveguide coupling from the sensing region has multiple benefits as it provides narrow fringes with high sensitivity that are independent of the length of the sensing region. This is achieved by allowing for large enough distances between the output ports. For example, the interferometer in Fig.~\ref{fig:Int} d) will experience the crosstalk in the sensing region, while that in Fig.~\ref{fig:Int} e) is less likely due to the negligible next-to-nearest-neighbour coupling. The elimination of the crosstalk is the main limiting factor in reduction of the interferometer footprint. The minimum distance between the output ports that ensures the crosstalk-free light propagation in the sensing region directly sets the minimum coupler width. For the waveguides with low refractive-index contrast, such as those written by direct laser writing in glass, typical minimum distances can be conservatively approximated by 100 $\mu$m \cite{GraefeThesis2017}. For the waveguide with high refractive-index contrast, for example those in silica-on-insulator (SOI), the minimum distances are down to tens of nanometers \cite{JaneiroSciRep2019, VelhaOptComms2019}. To appreciate the small footprint of the PL-based interferometers, we compare these values with the minimum distances required in directional couplers. Therein, the minimum distance between ports is given by the minimum bend radius, which is, for the S-bends, of the order of 15 mm in glass and 600 $\mu$m in SOI~\cite{PeruzzoScience2010}. This indicates that the footprint of PL interferometers can be reduced by 2 orders of magnitude with respect to the standard interferometers based on directional couplers. The planar architecture of the proposed interferometers not only occupies a significantly smaller space, but also enables stacking into a multilayer circuit or a 2D lattice.

In a different approach, the coupling between waveguides in the sensing region can be controlled to  achieve correlations between the interferometer modes. The recent free-space experiments have shown that the entanglement between the interferometer modes works towards the Heisenberg sensitivity enhancement \cite{LiuNatPhoton2021}. The proposed PL-based couplers support the propagation of a single or multiple photons by the continuous quantum walk and as such are a new playground for development of the distributive quantum measurements. However, the depth and size of this problem require a separate analysis that lies out of scope of this paper.

Multiarm interferometry opens up the possibilities for implementation of differential and multiparameter sensing schemes. In a differential measurement, subtraction of signal from the reference recorded at a different output port and weighted if necessary, Fig.~\ref{fig:Scheme}, allows for elimination of the intensity fluctuations arising from the instabilities in the source and environmental conditions. In a multiparameter measurement, simultaneous measurements at $M$ output ports provide information sufficient to retrieve $M$ different parameters. For example, interferometers in Fig.~\ref{fig:Int} c) and f) are suitable for 2-parameter measurement at the ports 2 and 4 or the ports 1 and 5, respectively. Moreover, the multiparameter measurement is often a method of choice for the elimination of cross-sensitivity, a commonplace problem in sensing which occurs when the interferometer response to the measured parameter depends on the parameters that are parasitic to the measurement \cite{BhatiaOptLett1997}. Finally, by allowing for waveguide coupling in the sensing region, the proposed interferometers can become tools for the multiphase quantum sensitivity estimation ~\cite{GessnerPRL2018, CiampiniSciRep2021}.

We briefly comment on the impact of the unwanted long-range coupling between waveguides. It perturbs the light propagation thus affecting the output balance of $1\times N$ splitters and the sensitivity of interferometers proposed here. However, the long-range coupling does not compromise the commensurability principle, leaving it possible to construct equal splitters by inclusion of the higher order diagonals in equation (\ref{eq:H}).

\section{Conclusion}\label{sec:Conclusion}
We have proposed a new type of multiarm optical interferometers on chip, which shows the maximum possible sensitivity enhancement and has convenient planar architecture with a small footprint. These advances over the standard directional-coupler architectures are achieved by the layout based solely on the modulated finite photonic lattices with commensurable eigenvalues. The nontrivial inverse design of the equal-power splitters benefits from the periodic light propagation through these lattices and permits the analytical solutions. While the crosstalk-free sensing region ensures the maximum sensitivity of a classical measurement, the proposed interferometers can be easily accommodated for the use in quantum distributive measurements with correlated modes. A particularly interesting future prospect is their application in classical and quantum multiparameter estimation.

\section{Acknowledgments}
J. Petrovic acknowledges support by the Berliner ChancengleichheitsProgramm (BCP). She did part of the work at the Institut f\"{u}r Optik und Atomare Physik, Technische Universit\"{a}t Berlin, Germany. A. Maluckov acknowledges  support  by  the  Ministry  of Education,  Science,  and  Technological  Development  of  the Republic  of  Serbia,  Grant  No.  451-03-68/2022-14/200017.




\begin{thebibliography}{99}
\bibitem{GiovannettiScience2004} V. Giovannetti, S. Lloyd and L. Maccone, Quantum-Enhanced Measurements: Beating the Standard Quantum Limit, Science, 306  (2004) 1330-1336.
\bibitem{WeitzPRL1996} M. Weitz, T. Heupel and T. W. H\"{a}nsch, Multiple Beam Atomic Interferometer, Phys. Rev. Lett., 77 (1996) 2356-2359.
\bibitem{PetrovicNJP2013} J. Petrovic, I. Herrera, P. Lombardi, F. Sch\"{a}fer, F. S. Cataliotti, A Multi-State Interferometer on an Atom Chip, New J. Phys., 15 (2013) 043002
\bibitem{SuPRL2017} Z.-E. Su, Y. Li., P. P. Rohde, H.-L. Huang,  X.-L. Wang, L. Li,N.-L.Liu, J. P. Dowling, C.-Y. Lu,  J.-W. Pan, Multiphoton Interference in Quantum Fourier Transform Circuits and Applications to Quantum Metrology, Phys. Rev. Lett., 119 (2017) 080502
\bibitem{PezzeNatPhoton2021} L. Pezz\'{e}, Entanglement-enhanced sensor networks, Nat. Photon., 15 (2021) 74-76.
\bibitem{ChennanOptExpress2016} C. Hu, Z. Yu and A. Wang, An all fiber-optic multi-parameter structure health monitoring system, Opt. Express, 24 (2016) 20287-20296
\bibitem{SunJRSM2020} L. Sun, M. Joshi, S.N. Khan, H. Ashrafian and A. Darzi, Clinical impact of multi-parameter continuous non-invasive monitoring in hospital wards: a systematic review and meta-analysis, J. Royal Soc. Med., 113 (2020) 217-224
\bibitem{SuzukiOptExress2006} K. Suzuki, V. Sharma, J. G. Fujimoto, E. P. Ippen, Y. Nasu, Characterization of symmetric [3 × 3] directional couplers fabricated by direct writing with a femtosecond laser oscillator, Opt. Express, 14 (2006), 2335-2343.
\bibitem{WeihsOptLett1996} G. Weihs, M. Reck, H. Weinfurter and A. Zeilinger, All-fiber three-path Mach–Zehnder interferometer, Opt. Lett., 21 (1996) 302-304
\bibitem{ChaboyerSciRep2015} Z. Chaboyer, T. Meany, L.G. Helt, M. J. Withford, M. J. Steel , Tunable quantum interference in a 3D integrated circuit, Sci. Rep., 5 (2015) 9601.
\bibitem{KeilAPLPhoton2016} R. Keil, T. Kaufmann, T. Kauten, S. Gstir, C. Dittel, R.
Heilmann, A. Szameit, G. Weihs, Hybrid waveguide-bulk multi-path
interferometer with switchable amplitude and phase, APL Photonics, 1 (2016) 081302.
\bibitem{SpagnoloSciRep2012} N. Spagnolo, L. Aparo, C. Vitelli, A. Crespi, R. Ramponi, R. Osellame, P. Mataloni and F. Sciarrino, Quantum interferometry with three-dimensional geometry, Sci. Rep., 2 (2012) 862.
\bibitem{MarcatiliBellSysTechJ1969} E. A. J. Marcatili, Bends in Optical Dielectric Guides, Bell Sys. Techn. J., 48 (1969) 2103–2132.
\bibitem{PolinoOptica2019} E. Polino, M. Riva, M. Valeri, R. Silvestri, G. Corrielli, A.
Crespi, N. Spagnolo, R. Osellame, F. Sciarrino, Experimental multiphase estimation on a chip, Optica 6 (2019) 288-295.
\bibitem{BrombergPRL2009} Y. Bromberg, Y. Lahini, R. Morandotti, Y. Silberberg, Quantum and classical correlations in waveguide lattices, Phys. Rev. Lett., 102 (2009) 253904.
\bibitem{BellecOptLett2012} M. Bellec, G. M. Nikolopoulos, S. Tzortzakis, Faithful communication Hamiltonian in photonic lattices, Opt. Lett., 37 (2012) 4504-4506.
\bibitem{GuzmanPRL2021} D. Guzm\'an-Silva, G. C\'aceres-Aravena and R. A. Vicencio, Experimental Observation of Interorbital Coupling, Phys. Rev. Lett., 127 (2021) 066601
\bibitem{PetrovicAnnPhys2018} J. Petrovic, J. J. P. Veerman, A new method for multi-bit and qudit transfer based on commensurate waveguide arrays,  Ann. Phys., 392 (2018) 128-141.
\bibitem{Perez-LeijaPRA2013} A. Perez-Leija, J. C. Hernandez-Herrejon, H. Moya-Cessa, A. Szameit, C. N. Demetrios, Generating photon-encoded $W$ states in multiport waveguide-array systems, Phys. Rev. A, 87 (2013) 013842.
\bibitem{PaulJOpt2014} S. Paul, K. Thyagarajan, Generation of N-partite single and two photon W states with enhanced tolerance using waveguide arrays, J. Opt. 16 (2014) 105503.
\bibitem{GraefeNatPhoton2014} M. Gr\"{a}fe, R. Heilmann, A. Perez-Leija, R. Keil, F. Dreisow,
M. Heinrich, H. Moya-Cessa, S. Nolte, D. N. Christodoulides, A. Szameit, On-chip generation of high-order single-photon W-states, Nat. Photon., 8 (2014) 791.
\bibitem{PetrovicOptLett2015} J. Petrovic, Multiport waveguide couplers with periodic energy exchange, Opt. Lett., 40 (2015) 139-142.
\bibitem{LedererPRL1999} T. Pertsch, P. Dannberg, W. Elflein, A. Bräuer, and F. Lederer, Optical Bloch Oscillations in Temperature Tuned Waveguide Arrays, Phys. Rev. Lett., 83 (1999) 4752.
\bibitem{PetrovicAO2008} J. Petrovic, Y. Lai, and I. Bennion, Numerical and experimental study of microfluidic devices in step-index optical fibers, Appl. Opt., 47 (2008) 1410-1416.
\bibitem{CrespiLOC2010} A. Crespi, Y. Gu, B. Ngamsom, H. J. W. M. Hoekstra, C. Dongre, M. Pollnau, R. Ramponi,  H. H. van den Vlekkert, P. Watts, G. Cerullo, R. Osellame, Three-dimensional Mach-Zehnder interferometer in a microfluidic chip for spatially-resolved label-free detection, Lab Chip, 10 (2010) 1167-1173.
\bibitem{ZhangOptica2021} M. Zhang, C. Wang, P. Kharel, D. Zhu, M. Lon\v{c}ar, Integrated lithium niobate electro-optic modulators: when performance meets scalability, Optica, 8 (2021) 652-667.
\bibitem{GraefeThesis2017} M. Gr\"{a}fe, Integrated Photonic Quantum Walks in Complex Lattice Structures, PhD Thesis, 2017.
\bibitem{JaneiroSciRep2019} Janeiro, R., Flores, R., Viegas, J. Silicon photonics waveguide array sensor for selective detection of VOCs at room temperature. Sci. Rep., 9 (2019) 17099.
\bibitem{VelhaOptComms2019} P. Velha, I. Cerutti, N. Andriolli, Crosstalk and BER performance of closely-spaced silicon-on-insulator waveguide arrays, Opt. Commun., 437 (2019) 214-218.
\bibitem{PeruzzoScience2010} A. Peruzzo, M. Lobino, J. C. F. Matthews, N. Matsuda, A. Politi, K. Poulios, X.-Q. Zhou, Y. Lahini, N. Ismail, K. W\"{o}rhoff, Y. Bromberg, Y. Silberberg, M. G. Thompson, Jeremy L. O'Brien. Quantum Walks of Correlated Photons, Science, 329 (2010)  1500-1503.
\bibitem{LiuNatPhoton2021} L.-Z. Liu, Y.-Z. Zhang, Z.-D. Li, R. Zhang, X.-F. Yin, Y.-Y. Fei, L. Li, N.-L. Liu, F. Xu, Y.-A. Chen, J.-W. Pan, Distributed quantum phase estimation with entangled photons, Nat. Photon., 15 (2021) 137.
\bibitem{BhatiaOptLett1997} V. Bhatia, D. Campbell, R. O. Claus, A. M. Vengsarkar, Simultaneous strain and temperature measurement with long-period gratings, Opt. Lett., 22 (1997) 648-650.
\bibitem{GessnerPRL2018} M. Gessner, L. Pezz\'{e}, A. Smerzi, Sensitivity Bounds for Multiparameter Quantum Metrology, Phys. Rev. Lett., 121 (2018) 130503.
\bibitem{CiampiniSciRep2021} M. A. Ciampini, N. Spagnolo, C. Vitelli, L. Pezz\`{e}, A. Smerzi, F.
Sciarrino, Quantum-enhanced multiparameter estimation in multiarm interferometers, Sci. Rep., 6 (2016) 28881.
\end{thebibliography}
\end{document}